\documentclass[aps, prl, twocolumn, superscriptaddress]{revtex4-2}
\usepackage{graphicx}
\usepackage{booktabs}
\usepackage{CJKutf8, color}
\usepackage[dvipsnames]{xcolor} 
\usepackage{graphicx}\usepackage{subcaption}
\usepackage{ulem} 
\usepackage{soul} 
\usepackage{appendix}
\newcommand{\aap}{Astron. \& Astrophys.}
\newcommand{\mnras}{Mon. Not. R. Astron. Soc.}
\newcommand{\apjl}{Astrophys. J. Lett.}
\newcommand{\apjs}{Astrophys. J. Suppl.}
\newcommand{\apss}{Astrophys. Spa. Sci.}
\newcommand{\nphysa}{Nucl. Phys. A}
\newcommand{\pasj}{Publ. Astron. Soc. J.}

\newcommand{\HU}{Graduate School of Advanced Science and Engineering, Hiroshima University, 1-3-1 Kagamiyama, Higashi-Hiroshima, Hiroshima 739-8526, Japan}

\newcommand{\TUS}{Department of Physics, Tokyo University of Science, 1-3 Kagurazaka, Shinjuku, Tokyo 162-8601, Japan}
\newcommand{\RIKENPRI}{RIKEN Pioneering Research Institute (PRI), 2-1 Hirosawa, Wako, Saitama 351-0198, Japan}

\newcommand{\iTHEMS}{RIKEN Center for Interdisciplinary Theoretical \& Mathematical Sciences (iTHEMS), RIKEN 2-1 Hirosawa, Wako, Saitama 351-0198, Japan}

\newcommand{\CNS}{Center for Nuclear Study (CNS), The University of Tokyo, 7-3-1 Hongo, Bunkyo, Tokyo 113-0033, Japan}

\newcommand{\IMP}{Institute of Modern Physics, Chinese Academy of Sciences, Lanzhou, Gansu 730000, China}
\newcommand{\CAS}{School of Nuclear Science and Technology, University of Chinese Academy of Sciences, Beijing 100049, China}
\newcommand{\Kogakuin}{Academi Support Center, Kogakuin University, 65-1 Nakano-machi, Hachioji, Tokyo 192-0015, Japan}

\newcommand{\kanji}[1]{\begin{CJK}{UTF8}{ipxm}(#1)\end{CJK}}

\newcommand{\chinese}[1]{\begin{CJK}{UTF8}{gbsn}(#1)\end{CJK}}

\begin{document}

\begin{CJK*}{}{}

\title{Evidence of Nuclear Urca Process in the Ocean of \\Neutron-Star Superburst MAXI~J1752$-$457}

\author{\mbox{Hao~Huang~\chinese{黄皓}}}
\email[]{hao.huang@a.riken.jp}
\affiliation{\RIKENPRI}
\affiliation{\IMP}
\affiliation{\CAS}

\author{\mbox{Akira~Dohi~\kanji{土肥明}}}
\email[]{akira.dohi@riken.jp}
\affiliation{\RIKENPRI}
\affiliation{\iTHEMS}

\author{\mbox{Amira~Aoyama~\kanji{青山有未来}}}
\email[]{amira.aoyama@riken.jp}
\affiliation{\TUS}
\affiliation{\RIKENPRI}

\author{\mbox{Tomoshi~Takeda~\kanji{武田朋志}}} 
\email[]{ tomoshi.takeda@a.riken.jp}
\affiliation{\HU}
\affiliation{\RIKENPRI}

\author{\mbox{Nobuya~Nishimura~\kanji{西村信哉}}}
\email[]{nobuya@cc.kogakuin.ac.jp}
\affiliation{\Kogakuin}
\affiliation{\CNS}
\affiliation{\RIKENPRI}

\date{\today}

\begin{abstract}
We propose that the rapid cooling of the neutron star following its X-ray superburst in MAXI J1752$-$457 over a period of 4 days, observed by two Japanese satellites, MAXI and NinjaSat, is due to enhanced neutrino emission from cycles of electron capture and $\beta^{-}$ decay involving odd-$A$ nuclei (or Urca pairs) in the ocean. Hence, this work provides the first indication of the possible existence of such a ``nuclear Urca process". The observation of MAXI J1752$-$457 implies a hot ignition layer with a maximum temperature of $\sim4~{\rm GK}$, located near the Urca shell in the ocean, such that the nuclear Urca process becomes dominant for up to $\sim2$ days after the superburst. This behavior is distinct from that of normal Type-I X-ray bursts, which are triggered by hydrogen or helium burning in much shallower layers than those of superbursts. Our findings enable probing of superburst ashes through Urca pairs via long-term monitoring of crust cooling on day-long timescales.
\end{abstract}


\maketitle
\end{CJK*}




Type I X-ray bursts (hereafter X-ray bursts), which are X-ray transients driven by nuclear burning on accreting neutron stars (NSs), play an important role in the production of heavy nuclei. The long tails of the burst light curves, whose duration is $\sim 1$--$100~\mathrm{s}$, indicate the occurrence of nuclear burning on the proton-rich side of the nuclear chart, including the rp-process \cite{1981ApJS...45..389W, 2001PhRvL..86.3471S}. Among the $\sim120$ observed X-ray bursters, few events exhibit significantly longer durations of $\sim 10^3~\mathrm{s}$. These events are categorized as \textit{superbursts}, driven by carbon fusion, whereas regular X-ray bursts are triggered by hydrogen or helium ignition. Nucleosynthesis associated with the tails of these light curves has been extensively investigated for both regular bursts and superbursts. Additionally, subsequent production of neutron-rich nuclei in the NS crust during the cooling phase has been suggested \cite{2007ApJ...662.1188G, 2018ApJ...859...62L, 2021MNRAS.500.2958M}.


The behavior of X-ray bursts is governed by the competition between nuclear burning in the accretion layer and cooling of the NS. At sufficiently late times following the burst, the latter cooling effect becomes dominant. In general, the thermal flux after an X-ray burst, $F(t)$, can be described by two power-law components as
\begin{eqnarray}\label{eq:ft}
F(t)\propto \left\{
\begin{array}{c}
 (t-t_0)^{-\alpha_\mathrm{f}}    \quad \mathrm{for~} t< t_{\rm break}\,,\\
(t-t_0)^{-\alpha_\mathrm{s}}    \quad \mathrm{for~} t\ge t_{\rm break}\,, \\
\end{array}
\right.
\end{eqnarray}
where $t_0$ is the onset time of the superburst, $t_{\rm break}$ is the break time corresponding to a change in the slope, and $\alpha_\mathrm{f}$ and $\alpha_\mathrm{s}$ denote the power-law indices during the \textit{first} and \textit{second} cooling epochs, respectively. Although the values $\alpha_\mathrm{f} = 0.2$ and $\alpha_\mathrm{s} = 4/3$ \cite{2004ApJ...603L..37C} are widely accepted and used as the standard ones, they depend on NS microphysics such as the specific heat and thermal conductivity \cite{2006ApJ...646..429C}.

Regarding $\alpha_\mathrm{s}$, this index has been measured for many X-ray burst sources, and the observed variations could be attributed mainly to differences in the specific heat and ignition depth \cite{2009A&A...497..469I, 2011MNRAS.413.1913Z, 2014A&A...562A..16I, 2017A&A...604A..77K}. In contrast, there are few definitive measurements of $\alpha_\mathrm{f}$. From a theoretical perspective, because the temperature profile is highly non-uniform during the initial cooling phase after X-ray bursts, uncertainties in the microphysics affecting $\alpha_\mathrm{f}$ are quite complex \cite{2021MNRAS.500.4491Y}. In particular, this complexity becomes more pronounced in the case of superbursts, because the thermal diffusion timescale and characteristic temperatures are higher than those of normal X-ray bursts \cite{2006ApJ...646..429C, 2012ApJ...752..150K}. Coupled with the paucity of \textit{long-term} superburst observations (see also a recent catalog \cite{2023MNRAS.521.3608A}), extracting NS physics from superbursts remains challenging, although attempts based on recurrence times have been made using numerical models \cite{2003ApJ...595.1077C, 2006ApJ...646..429C, 2022ApJ...937..124D, 2024MNRAS.535.1575M}.

Recently, two Japanese X-ray satellites, MAXI and NinjaSat, have observed thermal cooling following the superburst of the new X-ray source MAXI J1752$-$457 \cite{2025ApJ...986L..29A}. The estimated indices in Eq.~(\ref{eq:ft}) are $\alpha_\mathrm{f} = 0.9\pm0.2$ and $\alpha_\mathrm{s} = 1.7\pm0.7$, implying faster early-time cooling compared to the standard case \cite{2004ApJ...603L..37C}. Thus, the cooling source that operates below the carbon ignition pressure ($P_{\rm ign}\sim10^{26}~{\rm dyn~cm^{-2}}$) should be more effective. As a plausible candidate for explaining such rapid cooling behavior, we propose cycles of electron capture and its inverse, $\beta^{-}$ decay via odd-$A$ nuclei in the NS ocean, i.e., "nuclear Urca process".

The importance of the nuclear Urca process in accreting NSs was first pointed out by Ref.~\cite{2014Natur.505...62S} and further investigated by Ref.~\cite{2015ApJ...809L..31D}, who identified 85 possible Urca pairs. The effects of Urca cooling on the thermal evolution of accreting NSs have been studied in detail, including studies focusing on MAXI J0556$-$332 \cite{2015ApJ...809L..31D, 2016ApJ...831...13D} and on carbon ignition triggering the superburst \cite{2024MNRAS.535.1575M}. In particular, the ${}^{63}{\rm Fe}$--${}^{63}{\rm Mn}$ Urca pair could significantly affect crust cooling in soft X-ray transients, depending on the properties of their nuclear structure \cite{2018JPhG...45i3001M, 2025ApJHao}. However, the occurrence of the Urca process has not yet been confirmed due to several physical uncertainties, in particular for the extra crust heating sources implied from observed hot X-ray transients (shallow heating) \cite{2009ApJ...698.1020B}. In this \textit{Letter}, we show that fast early-phase cooling (a higher $\alpha_\mathrm{f}$ than the standard value) in MAXI J1752$-$457 provides the first evidence of the operation of the nuclear Urca process in the ocean.


We here estimate the timescale of the Urca mechanism in the physical environment of the NS ocean. Assuming that the electron contribution dominates the specific heat $c_p$ and that electron-ion scattering determines the thermal conductivity $K$, following Ref.~\cite{2004ApJ...603L..37C}, the diffusion timescale associated with neutrino Urca cooling at a depth $H$ can be written as
\begin{eqnarray}
t_{\rm diff} &=& 
\frac{H^2\rho c_p}{K}
\nonumber \\ 
&\simeq& 2.0~{\rm days}\;
\left( \frac{H}{100~{\rm m}} \right)^{2}
\left( \frac{\rho}{10^{8}~{\rm g\,cm^{-3}}} \right)^{1/3}
 \nonumber \\
&&\times\left( \frac{Y_{\rm e}}{0.5} \right)^{-2/3}\left(\frac{\Lambda_{\rm ei}}{1} \right)\left(\frac{Z}{26} \right)^2
\left(\frac{A}{56}\right)^{-1}~,
\label{eq:tdiff}
\end{eqnarray}
where $\rho$ is the mass density, $Z$ is the atomic number, $Y_{\rm e}=Z/A$ is the electron fraction, and $\Lambda_{\rm ei}$ is the Coulomb logarithm for electron-ion scattering. This estimate implies that ocean Urca cooling operates primarily over the first few days and therefore primarily affects the early cooling phase of MAXI~J1752$-$457.

What temperature ranges can the Urca process influence in NS crust-cooling curves? The neutrino luminosity of Urca pairs is given by \cite{1970Ap&SS...7..374T,2016ApJ...831...13D}:
\begin{eqnarray}
    L_{\nu}\approx \left(\sum_{i}X_i\cdot L_{34,i}\right)\times10^{34}~{\rm erg~s^{-1}}~T_9^5\left(\frac{2}{g_{14}}\right)R_{10}^2~, \label{eq:Lnu}
\end{eqnarray}
where $T_9$ is the temperature in units of $10^9~{\rm K}$, $g_{14}$ is the surface gravity in units of $10^{14}{\rm cm~s^{-2}}$, and $R_{10}$ is the stellar radius in units of $10~{\rm km}$. Here, $i$ labels individual Urca shells, $X_i$ is the mass fraction of the corresponding Urca pair, and $L_{34,i}$ is a function of nuclear properties only \cite{2016ApJ...831...13D}:
\begin{eqnarray}
    L_{34,i}=0.87\left(\frac{10^6\,\mathrm{s}}{ft}\right)\left(\frac{56}{A}\right)\left(\frac{Q_{\mathrm{EC}}}{4\,\mathrm{MeV}}\right)^5\left(\frac{\langle F\rangle^*}{0.5}\right)~, \label{eq:L34}
\end{eqnarray}
where the $ft$-values measure the transition strength of the weak interactions and the $Q_{\rm EC}$ is the absolute value of the electron capture energy threshold. $\langle F\rangle^*=\langle F\rangle^+\langle F\rangle^-/(\langle F\rangle^++\langle F\rangle^-)$, where $\langle F\rangle^\pm\approx2\pi\alpha Z/|1-{\rm exp}(\mp2\pi\alpha Z)|$ is the Coulomb factor. Urca cooling can dominate the late-time thermal evolution following a superburst when $L_\nu \gtrsim 10^{37}{\rm ergs^{-1}}$, corresponding to roughly $10\%$ of the Eddington luminosity. This condition can be rewritten as
\begin{eqnarray}
    T_9 \gtrsim 3.47\left(\frac{\sum_{i}X_i\cdot L_{34,i}}{2}\right)^{-1/5}\left(\frac{g_{14}}{2}\right)^{2/5}
    .
\end{eqnarray}
Such high temperatures can be achieved in superbursters \cite{2004ApJ...603L..37C, 2023AstL...49..824K}, but not in normal X-ray bursters, where typical temperatures remain below $T_9 \lesssim 2.5$ (e.g., \cite{2004ApJ...603..242K, 2008ApJS..178..110P}). This implies that Urca cooling can affect the early crust-cooling phase following a superburst, as we demonstrate later.

Among the 85 candidate Urca pairs identified by \cite{2016ApJ...831...13D}, 15 pairs dominate the cooling in the NS ocean. These ocean Urca pairs are distributed over pressure ranges of $1.85\times10^{24} \le P/[{\rm dyn~cm^{-2}}] \le 2.2 \times 10^{27}$. 
A quantitative prediction for the neutrino luminosity requires the crustal composition and detailed nuclear-physics inputs, both of which remain substantial uncertainties. We therefore adopt a simplified phenomenological treatment and represent the ocean Urca cooling by two sources located at $\log_{10} \left(P/[{\rm dyn~cm^{-2}}]\right)=26$ and $26.5$, corresponding to depths of $H = 72$ and 96~m, respectively. 
The associated cooling strengths are set to $X_1 L_{34,1}=2.4$ and $X_2 L_{34,2}=0.8$, yielding a total Urca cooling strength of $\sum_i X_i L_{34,i}=3.2$ \footnote{To verify the assumption to treat only two Urca pairs, we also performed the superburst calculation with 15 Urca pairs in the ocean, suggested by \cite{2016ApJ...831...13D}. Concretely, we take the column depth(or pressure) and $L_{34,i}$ values of each Urca pair from \cite{2016ApJ...831...13D}, and their mass fraction from \cite{2012ApJ...752..150K}. In the case of $\Sigma_iX_iL_{34,i}=3.2$ realized by equally rescaling $L_{34,i}$ values, we confirm the same conclusion as in the case with two Urca pairs, i.e., the presence of Urca cycles is in better agreement with the observations of MAXI J1752$-$457.}. Ref.~\cite{2016ApJ...831...13D} estimated a value of $\sum_i X_i L_{34,i}=1.6$ for superburst ashes. However, this value carries an uncertainty of approximately a factor of two arising from nuclear-physics uncertainties, primarily due to experimental errors in the $\beta^+$-decay transition probabilities, characterized by the $ft$ values ($\propto L_\nu^{-1}$)~\footnote{By examining ENSDF data\cite{ENSDF}, we confirmed that the $3\sigma$ uncertainties in $\log(ft)$ for the ocean Urca pairs considered by~\cite{2016ApJ...831...13D} indeed correspond to an uncertainty of about a factor of two, as noted in their study.}.

\begin{figure}
    \centering
    \includegraphics[width=1.0\linewidth]{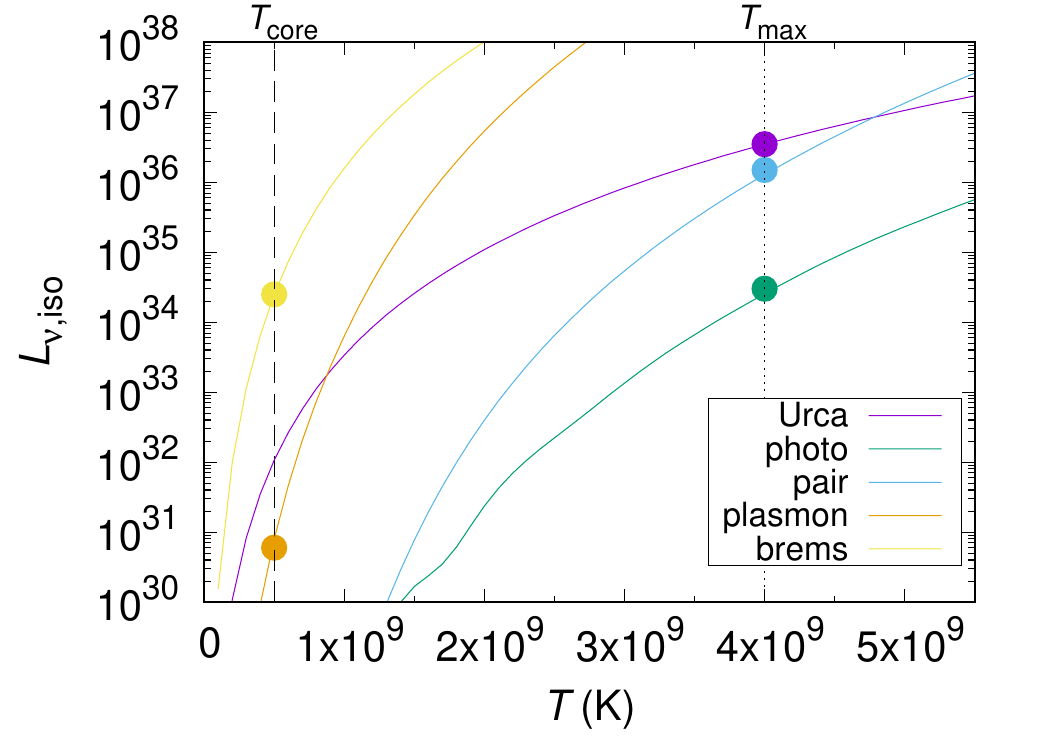}
    \caption{Neutrino luminosity of each process as a function of temperature. See texts for details.}
    \label{fig:Lnu}
\end{figure}

\begin{figure}
    \centering    
    \includegraphics[width=1.0\linewidth]{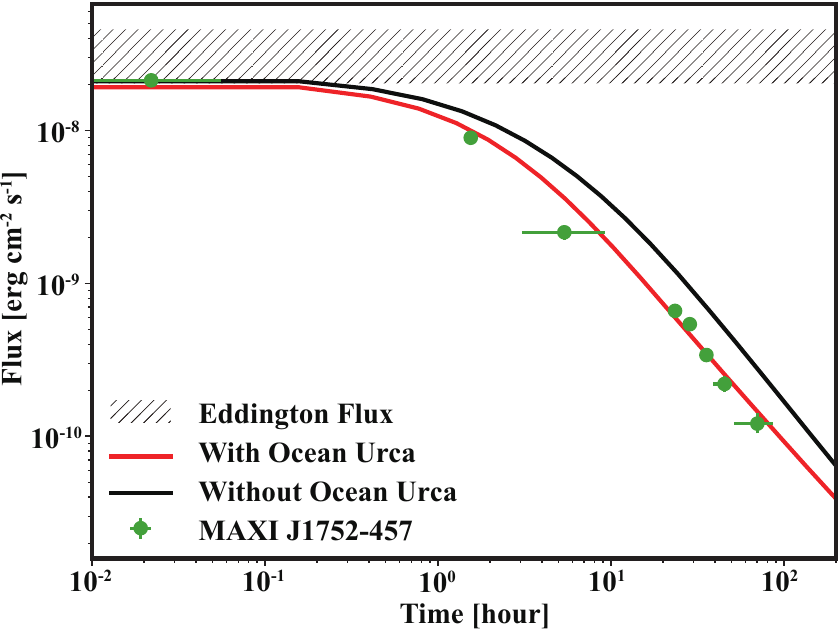}    
    \caption{Cooling curves without and with Urca cycles in the ocean, denoting black and red solid lines, for $M_{\rm core}=1.18 M_\odot$ and $R_{\rm core}=7.96~{\rm km}$ stars. The shadow region corresponds to the Eddington flux, including uncertainties of the hydrogen mass fraction.}
       \label{fig:lc}
\end{figure}

\begin{figure}
    \centering    
    \includegraphics[width=1.0\linewidth]{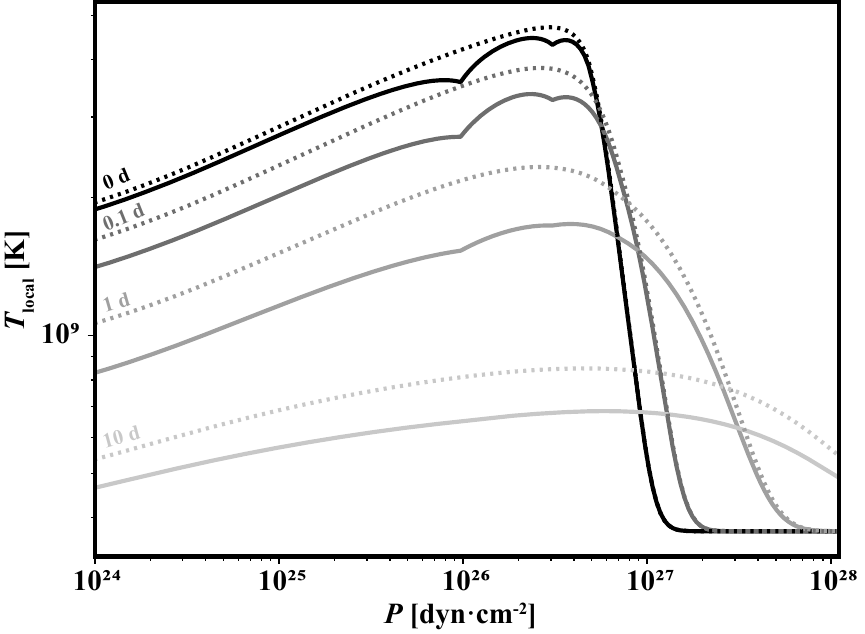}
    \caption{Evolution of temperature structure without and with Urca cycles in the ocean, denoting dashed and solid lines, respectively, for $M_{\rm core}=1.18 M_\odot$ and $R_{\rm core}=7.96~{\rm km}$ stars.
    }
    \label{fig:ty}
\end{figure}

\begin{figure*}
    \centering
    \begin{subfigure}[t]{0.48\textwidth}
        \centering        
        \includegraphics[width=\linewidth]{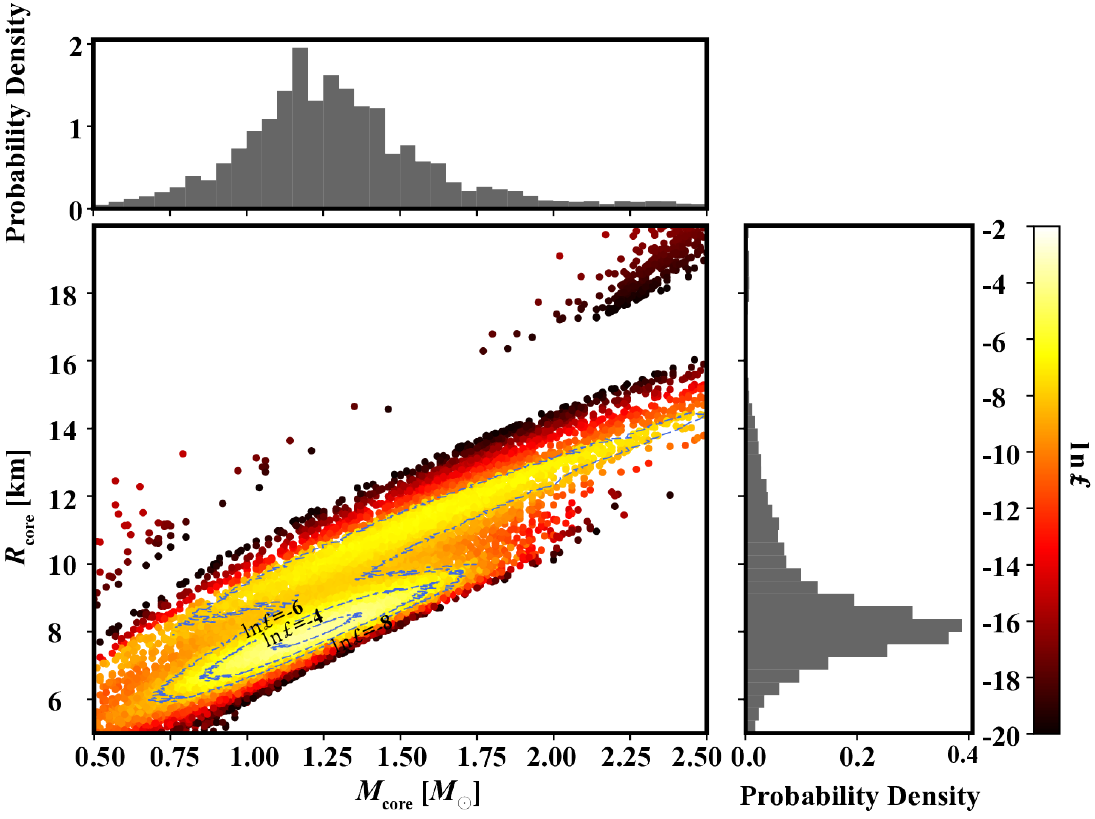}
        \caption{With Urca}
        \label{fig:pemr-left}
    \end{subfigure}\hfill
    \begin{subfigure}[t]{0.48\textwidth}
        \centering        
        \includegraphics[width=\linewidth]{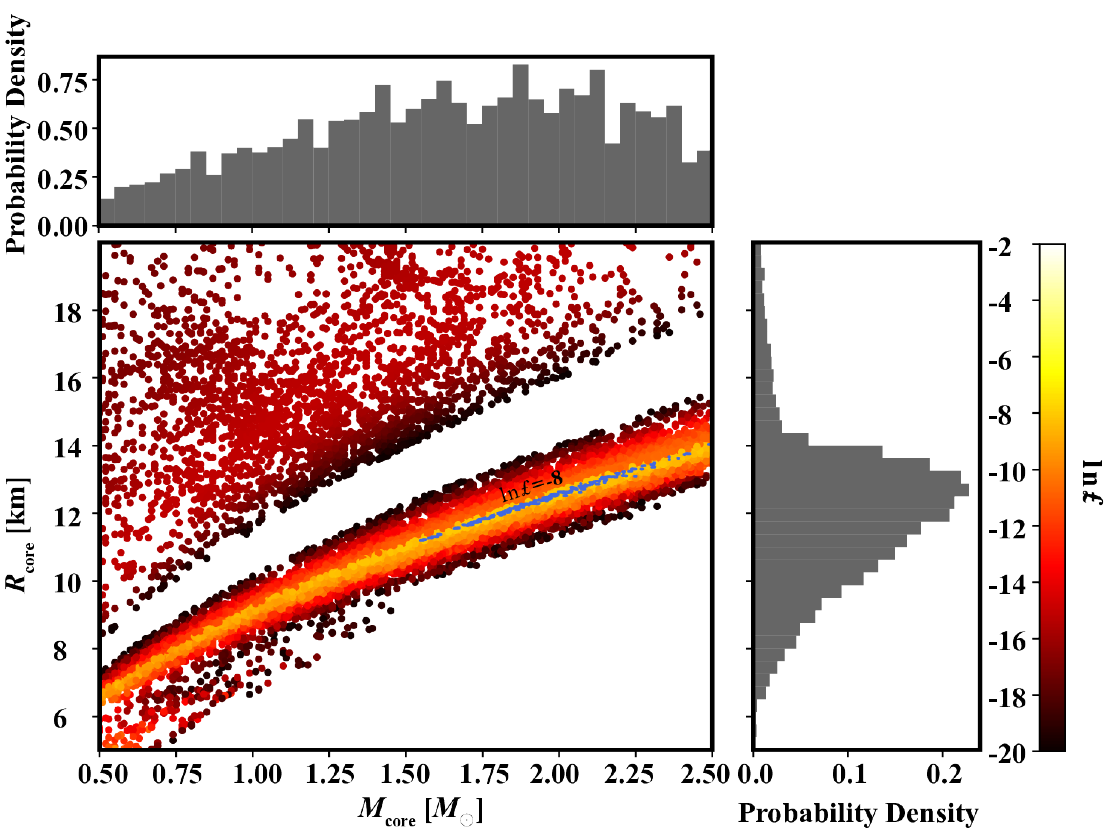}
        \caption{Without Urca}
        \label{fig:pemr-right}
    \end{subfigure}
    \caption{The log-likelihood distribution in the mass-radius plane obtained from the MCMC analysis. Blue contours corresponding to $\ln\mathcal{L}=-4$, -6, and -8 are shown, although no contours with $\ln\mathcal{L}\ge -6$ appear in panel (b). }
    \label{fig:pemr}
\end{figure*}

Following the same treatment as in \cite{2006ApJ...646..429C, 2014Natur.505...62S}, we estimate the neutrino luminosity in the accreting NS. Then, we assume the Helmholtz EOS by \cite{2000ApJS..126..501T} and pure ${}^{56}{\rm Fe}$ compositions, whose treatments are actually the same as those in our later cooling calculation. Figure~\ref{fig:Lnu} compares the Urca neutrino luminosity with other neutrino-emission processes, i.e., the photo-neutrino process, pair annihilation, plasmon decay, and bremsstrahlung \cite{1996ApJS..102..411I}, assuming an isothermal profile for a NS with mass $M_{\rm NS}=1.4~M_\odot$ and radius $R_{\rm NS}=10~{\rm km}$ \footnote{We have confirmed that the qualitative behavior shown in Figure~\ref{fig:Lnu} is insensitive to uncertainties in the neutron-star mass and radius.}. For $T_9 \gtrsim 1$, plasmon decay and bremsstrahlung dominate over the nuclear Urca process and the other channels. However, while these two processes occur mainly in the crust, the remaining three operate predominantly outside the crust. Consequently, if a significant temperature gradient exists between the crust and the overlying layers, the relative importance of the cooling channels can change.

In superburst scenarios, the ocean is the hottest region. Assuming a plausible temperature gradient for a superburster, such as an optimistic crust temperature of $T_{\rm core} = 5 \times 10^8~{\rm K}$\footnote{$T_{\rm core}$, which is the input parameter in \textit{dSTAR} code, is defined as the crust-core boundary temperature. Although actual NS crust is not isothermal, the temperature variation on the whole crust is not large (typically less than a factor of 2) \cite{2009ApJ...698.1020B,2014Natur.505...62S}, compared to the temperature difference between ocean and crust.}\footnote{Based on inferences from previous observations of accreting NSs, the maximum $T_{\rm core}$ is likely $\sim2\times10^8~{\rm K}$ (e.g., Fig.7 of \cite{2017PhRvC..95b5806C}).} and a maximum temperature near the ignition layer of $T_{\rm max}=4\times10^9~{\rm K}$, Urca cooling in the ocean can become dominant (filled circles in Figure~\ref{fig:Lnu}). As we show later, such conditions can be naturally realized during the early cooling phase of the superburst from MAXI~J1752$-$457.

To quantify the impact of the Urca process on MAXI~J1752$-$457, we construct thermal evolution models following the same methodology as \cite{2025ApJHao}. Specifically, we solve the general relativistic heat-diffusion equation using the \texttt{dSTAR} code \cite{2015ascl.soft05034B}. The microphysical inputs adopted in \texttt{dSTAR} follow Ref. \cite{2009A&A...497..469I}, except for the treatment of the $T_{\rm s}$-$T_{\rm b}$ relation and the implementation of the Urca process.

Following the same approach as in \cite{2004ApJ...603L..37C}, we assume that the superburst deposits energy uniformly in a surface layer $P \le P_{\rm ign}$, and that the energy released per unit mass of the ignited carbon, $E$, is uniformly distributed. To model the energy deposition in the NS crust during the superburst, we adopt an assumption of a 0.1-day accretion episode before the onset of cooling with accretion rate of $5\times10^{16}~{\rm g~s^{-1}}$. Although observations only weakly constrain the accretion history immediately before the superburst, the combination of the accretion duration and the heating strength determines the total released energy, which is well constrained observationally. After establishing a steady state at $t=0~{\rm d}$, we initiate the crust-cooling calculation by switching off the injected superburst energy. This phenomenological treatment has been widely used and validated for modeling the long tails of superbursts and long-duration X-ray bursts \cite{2023AstL...49..824K}.

We use the observational data of MAXI~J1752$-$457 with $1 \sigma$ uncertainties, assuming a fiducial distance of $D = 8 ~{\rm kpc}$ and no time lag between the onset of the superburst and the start of the observations, corresponding to $E_{17}=2.0$ and $y_{12}=2.8$ in right panel of Figure 4 in \cite{2025ApJ...986L..29A}\footnote{A time lag of up to 1.44hr is formally allowed by the observational cadence of MAXI; however, because the peak flux is limited by the Eddington luminosity, larger time lags are disfavored. Moreover, a longer time lag would imply faster early cooling, making scenarios without Urca cooling less likely}. Their observational errors are too small to alter our conclusion (see also Fig. 2 in \cite{2023AstL...49..824K}). 
In addition to the Urca-cycle properties, our model includes the neutron-star core mass and radius, $M_{\rm core}$ and $R_{\rm core}$, the ignition pressure $P_{\rm ign}$, and the heating strength $E$ as parameters in principle. We note that the crust impurity and the core temperature have negligible effects on the crust-cooling curves within the first $\sim$4~days after the superburst, because the heat flux during this phase is dominated by diffusion from the ocean \cite{2009ApJ...698.1020B, lalit2019, page2013}. Finally, since the observed flux of MAXI~J1752$-$457 is lower than that predicted by standard models, as discussed below, no additional heating source is required except the injected superburst energy.

We fit the calculated light-curve models with the observational data~\cite{2025ApJ...986L..29A} by employing a Markov Chain Monte Carlo (MCMC) approach, and we explore the parameter space of the neutron-star core mass and radius, $M_{\rm core}$ and $R_{\rm core}$, while other quantities, such as the heating strength $E=2.0 \times 10^{17}~{\rm erg\,g^{-1}}$ and the ignition pressure $P_{\rm ign}=5.0 \times 10^{26}~\mathrm{dyn~cm^{-2}}$\cite{2025ApJ...986L..29A}, are fixed. Broad, initially flat prior distributions are adopted: $0.5 \le M_{\rm core}/M_\odot \le 2.5$ and $5 \le R_{\rm core}/{\rm km} \le 20$. The parameter exploration is performed by coupling the \texttt{emcee} sampler \cite{foreman-mackey2013} with the \texttt{dSTAR} code.

In the \texttt{dSTAR} code, the surface temperature $T_{\rm s}$ and the bolometric flux are obtained using a $T_{\rm s}$--$T_{\rm b}$ relation, where $T_{\rm b}$ denotes the temperature at the outermost mesh point. The density at this outer boundary is fixed at $\rho_{\rm b} = 10^{6}~{\rm g~cm^{-3}}$. The $T_{\rm s}$-$T_{\rm b}$ relation originally implemented in \cite{2009ApJ...698.1020B} cannot reproduce the observations of MAXI~J1752$-$457 because of its high surface temperature, $T_{\rm s} \gtrsim 10^{6.7}~{\rm K}$ (see Fig. 2 of \cite{2025ApJ...986L..29A}). Instead, we adopt the $T_{\rm s}$--$T_{\rm b}$ relation presented by Fig. 1 in \cite{2020ApJ...888...97B}, which is applicable to high surface temperatures up to $T_{\rm s} \lesssim 10^{7.36}~{\rm K}$ (see their Fig.~1). We note that this $T_{\rm s}$--$T_{\rm b}$ relation approximately reproduces the extrapolation of the traditional ${}^{56}{\rm Fe}$ envelope model to higher $T_{\rm b}$ values \cite{1982ApJ...259L..19G}. For the burst ashes, we therefore assume a ${}^{56}{\rm Fe}$ envelope, consistent with the observational analysis of \cite{2025ApJ...986L..29A} and with theoretical expectations that Fe- and Ni-group nuclei dominate the composition of superburst ashes \cite{2003ApJ...583L..87S, 2011ApJ...743..189K}.


Figure~\ref{fig:lc} shows representative light curves for MAXI~J1752$-$457 with a NS mass of $1.18~M_\odot$, and a radius of $7.96~\mathrm{km}$, corresponding to the best-fit model that includes Urca cooling. The inclusion of ocean Urca cooling accelerates the post-superburst cooling and ultimately reproduces the observed cooling behavior. This result is also consistent with the observationally inferred larger first-decay index $\alpha_f$ compared to the standard theoretical expectation \cite{2004ApJ...603L..37C}. We therefore conclude that ocean Urca cooling is a viable mechanism contributing to the observed cooling of MAXI~J1752$-$457.

Our results differ quantitatively from those of Ref.~\cite{2016ApJ...831...13D}, who suggested only a weak effect of ocean Urca cooling on post-superburst light curves. At present, however, we do not identify a unique origin for this discrepancy. Because the Urca luminosity scales strongly with the local temperature, the difference may reflect details of the thermal evolution during the superburst and the subsequent cooling. Nevertheless, our conclusion remains based on the thermal condition realized in the present calculations for MAXI J1752-457 rather than on a specific interpretation of Ref. \cite{2016ApJ...831...13D}. 

The effect of ocean Urca cooling can be understood from the temperature evolution shown in Figure~\ref{fig:ty}. A valley-like structure is visible, corresponding to cooler regions around $P \sim 10^{26}~{\rm dyn~cm^{-2}}$ at $t \lesssim 1~{\rm d}$. This behavior is broadly consistent with the analytical diffusion timescale given by Eq.~(\ref{eq:tdiff}). During this early phase, immediately after the superburst, a strong temperature gradient develops: the explosive burning associated with the superburst—parameterized by $E_{17}$ and $P_{\rm ign}$—heats the ocean efficiently, while leaving the crust comparatively cool. At $t=0~{\rm d}$, the temperature near the hottest ocean layer reaches $\simeq 4~{\rm GK}$, whereas the crust temperature remains at $\simeq 0.4~{\rm GK}$. As discussed in Figure~\ref{fig:Lnu}, such a temperature contrast makes ocean Urca cooling dominant. This behavior arises from the close proximity of the Urca shell to the explosive burning layer produced by carbon fusion.

Figure~\ref{fig:pemr} therefore shows the likelihood distribution in the $M_{\rm core}$-$R_{\rm core}$ plane. For each set of $M_{\rm core}$ and $R_{\rm core}$, we perform a dStar calculation and compare the computed cooling curve with the observed light curve. Since $M_{\rm core}$ and $R_{\rm core}$ affect the surface gravity and gravitational redshift, they modify both the thermal evolution and the mapping to the observed flux. We notice that when Urca cooling is included, the maximum value of the log-likelihood reaches $\ln \mathcal{L} \simeq -3$. In contrast, in models without Urca cooling, the best-fit solution yields $\ln \mathcal{L} \simeq -8$, which fails to reproduce the observed light curve shown in Figure~\ref{fig:lc}. With Urca cooling, a well-defined region appears in the mass-radius plane where $\ln \mathcal{L} > -4$, indicating that the observations can be well reproduced for NS masses and radii within this region. No such region exists in the absence of Urca cooling, implying that models without Urca cannot adequately explain the observations. This result therefore provides supporting evidence for the presence of ocean Urca cooling. 

The inferred radius of $R \simeq 7.96~\mathrm{km}$ is smaller than other current observational constraints, which typically indicate $R \sim 10$-$14~\mathrm{km}$, as inferred from GW170817 and NICER observations (see, e.g., the recent review by \cite{MALIK2025100086}). This discrepancy may be alleviated by uncertainties in the source distance. If the actual distance is larger than the assumed value of $8~{\rm kpc}$, the intrinsic luminosity corresponding to the observed flux becomes higher. Under the simplifying assumption that the NS mass and radius enter the early cooling stage only through general-relativistic effects, the gravitational redshift factor $(1+z_g)$, which alters the flux normalization, must increase to reproduce the observed flux. For example, if future observations establish the distance to be as large as the current upper limit of $D = 12~\mathrm{kpc}$ \cite{2025ApJ...986L..29A}, the required $(1+z_g)$ decreases by a factor of $\sim1.5$, implying a larger stellar radius of $R \simeq 12.6~\mathrm{km}$ for a NS with $M = 1.16~M_\odot$. Nevertheless, we emphasize that uncertainties in the distance would not affect the steepness of the cooling curves. Consequently, our conclusion regarding the necessity of ocean Urca cooling remains robust against distance uncertainties.


From the above considerations, we propose that the rapid post-superburst decay observed in MAXI~J1752$-$457 provides the first evidence for the presence of Urca pairs in the NS ocean. This is because Urca shells can exist near the carbon ignition depth, $P_{\rm ign}\sim10^{26}~{\rm dyn~cm^{-2}}$, where sufficiently high temperatures are achieved for ocean Urca cooling to operate efficiently. In contrast, neutrino cooling processes in the crust remain subdominant owing to the much lower temperatures there, allowing ocean Urca cooling to dominate the early cooling phase. We emphasize that such conditions arise naturally in superbursts, whereas in normal X-ray bursts the ocean temperature is too low for Urca cooling to play a significant role.

Our findings open a new window to probe the origin of Urca pairs not only through theory, but also via nuclear-physics experiments and astronomical observations. Superburst nucleosynthesis indicates that odd-mass nuclei can be synthesized in the superburst ashes. For example, one-zone calculations \cite{2003NuPhA.718..247S} showed that about 8\% of the ashes consist of ${}^{57}{\rm Co}$ or ${}^{57}{\rm Fe}$, which can give rise to the ${}^{57}{\rm Fe}$--${}^{57}{\rm Mn}$ and ${}^{57}{\rm Mn}$--${}^{57}{\rm Cr}$ Urca pairs. More realistic multi-zone models \cite{2012ApJ...752..150K} demonstrated the production of ${}^{25}{\rm Mg}$ for pure-helium accretion and ${}^{69}{\rm As}$ for solar-composition accretion, which can serve as progenitors of the ${}^{25}{\rm Mg}$-${}^{25}{\rm Na}$ and ${}^{69}{\rm Zn}$-${}^{69}{\rm Cu}$ Urca pairs through three or four successive electron captures, respectively. More precise measurements of the relevant $ft$ values and of nuclear masses\footnote{Experimental uncertainties in nuclear masses affect electron-capture rates through the $Q$ values, and hence the strength of Urca cooling. While the relative uncertainties in $Q$ values for most electron-capture reactions relevant to ocean Urca pairs are at the few-percent level, those for some heavy nuclei, such as ${}^{79}{\rm As}$, ${}^{81}{\rm As}$, ${}^{101}{\rm Mo}$, and ${}^{101}{\rm Nb}$, can reach tens of percent according to AME2020 \cite{Huang2021AME2020PartII}.} will therefore be crucial for identifying the specific Urca pairs operating in MAXI~J1752$-$457 and in similar systems observed in the future.

Long-term observations of superbursters are essential for detecting the strong early cooling driven by the Urca process. At present, MAXI~J1752$-$457 is the longest-observed superburster (see also the available light-curve data in \cite{2023MNRAS.521.3608A}), owing to the excellent long-term monitoring capability of the CubeSat X-ray observatory NinjaSat \cite{2025PASJ...77..466T}. Because the extreme conditions realized in superbursts are particularly favorable for ocean Urca cooling, future X-ray observations combining prompt follow-up with extended monitoring will be crucial for testing this scenario—namely, whether other superbursters can also serve as laboratories for probing ocean Urca pairs. In current observations, the former role is fulfilled by MAXI, while the latter has been provided by NinjaSat, whose operation ended on September 17, 2025. Looking ahead, the planned launch of NinjaSat2 in 2028, designed to support both rapid response and long-term monitoring, will offer an important opportunity to further explore this phenomenon. For a discussion of alternative interpretations of the day-scale X-ray transient, including scenarios other than a superburst, see \cite{2025ApJ...986L..29A}.

We thank X.~Tang, K.~Li and Z.~Meisel for fruitful discussions. We also acknowledge support from the overseas program of RIKEN PRI (hosted by S.~Nagataki), which initiated this project. This work was supported by the National Key Research and Development Program (MOST 2022YFA1602304) and JSPS KAKENHI (Grant Nos. JP24H00008, JP25H01273, JP25K17403, JP25KJ0241, and JP26K00723).

\bibliography{ref}
\clearpage
\onecolumngrid
 
\begin{figure*}[t]
    \centering
   \begin{minipage}{0.49\linewidth}
       \includegraphics[width=\linewidth]{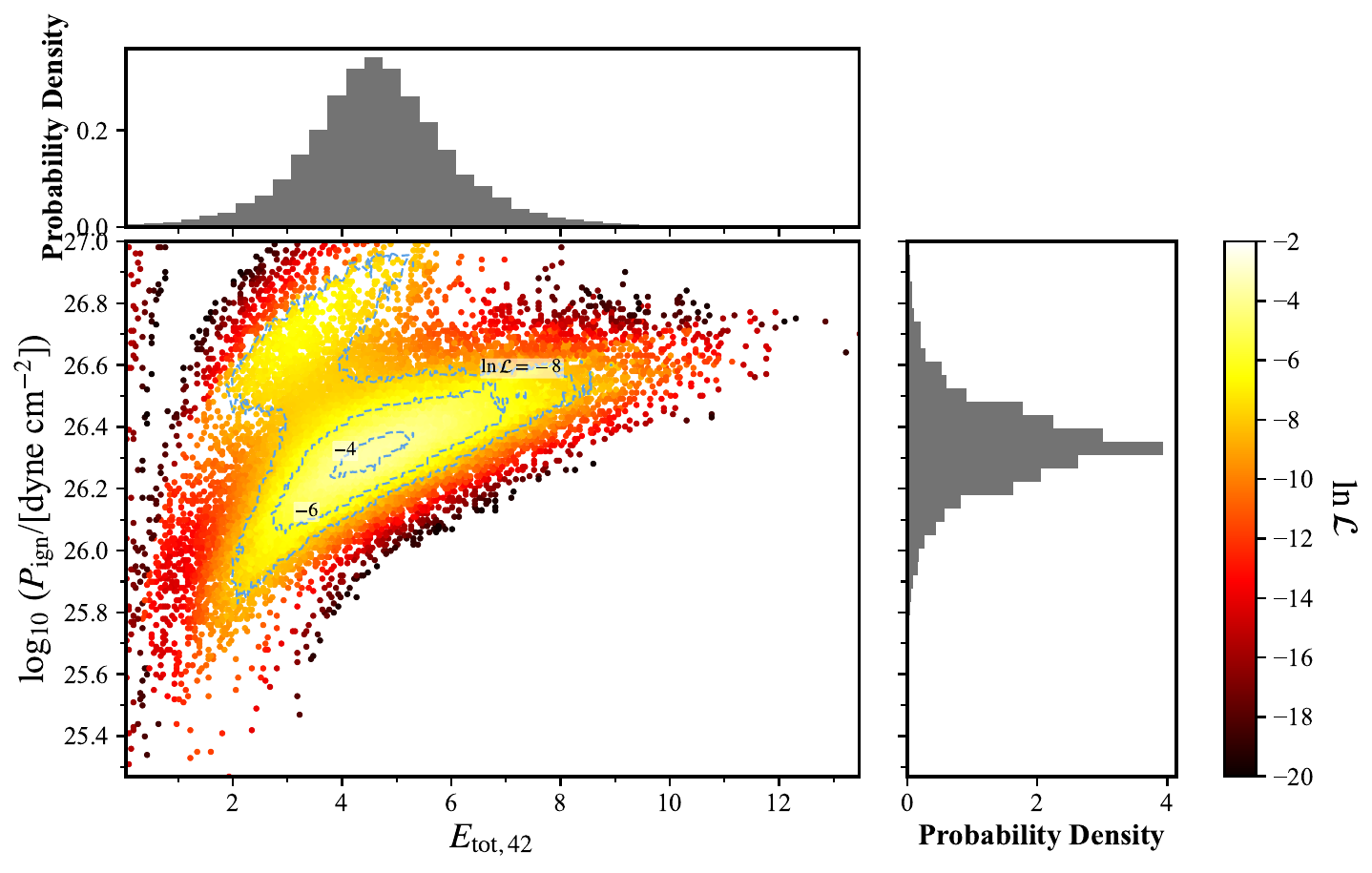}
   \end{minipage}
   \begin{minipage}{0.49\linewidth}
\includegraphics[width=\linewidth]{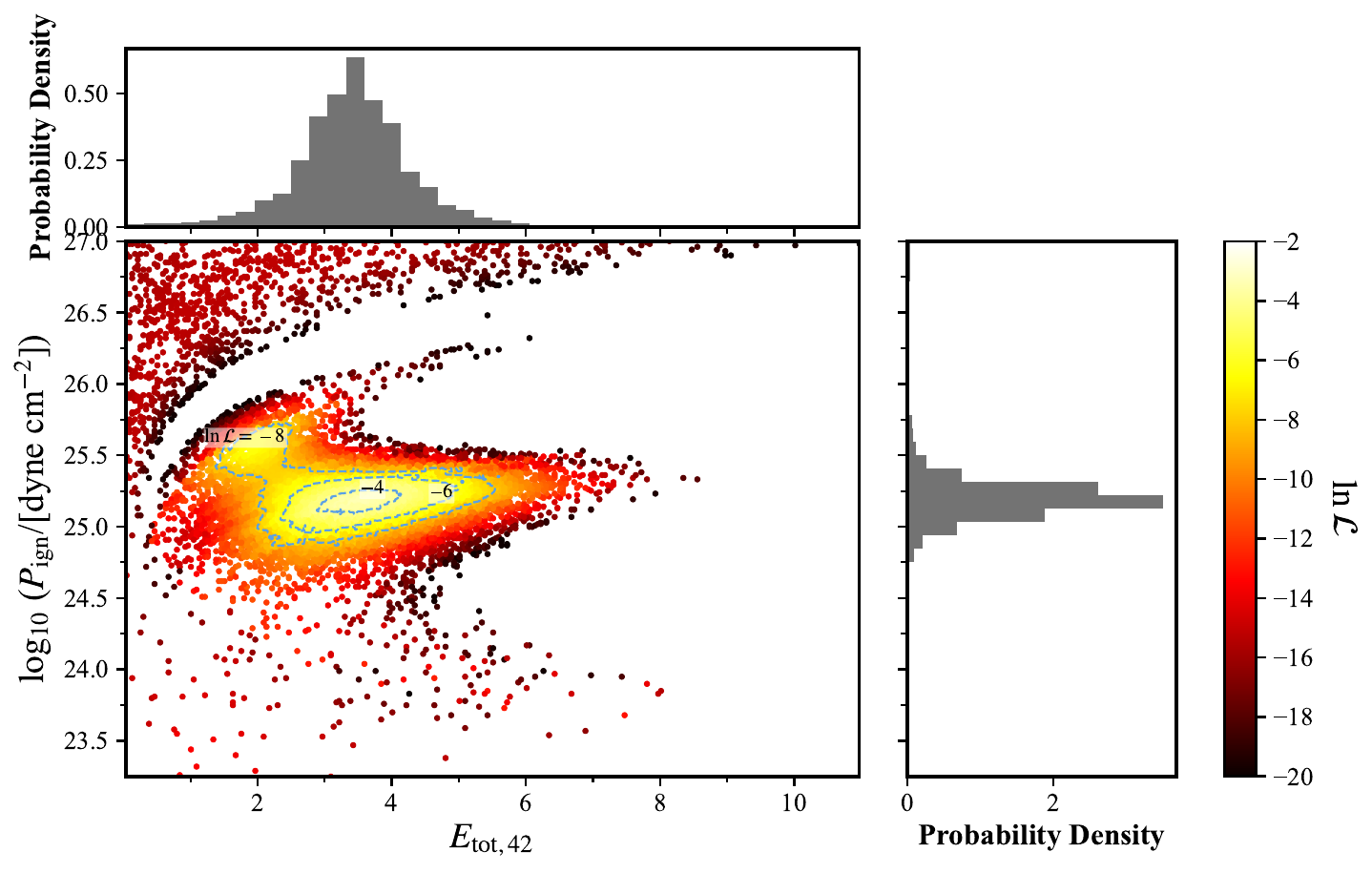}
\end{minipage}
    \caption{The log-likelihood distribution in the plane of ignition pressure and total burst energy in units of $10^{42}~{\rm erg}$ obtained from the MCMC analysis. The left panel includes Urca cooling, while the right panel does not.
}
    \label{fig:mcmcpe}
\end{figure*}

\section*{Appendix: On the MCMC test for consistency with $E_{17}$ and $P_{\rm ign}$}
Throughout this letter, we assumed the values of $E_{17}$ and $P_{\rm ign}$ as those reported by \cite{2025ApJ...986L..29A}. To verify this assumption, we test the MCMC calculation by changing $E_{17}$ and $P_{\rm ign}$ based on the same method in the letter. To save computational cost, we fix the neutron-star mass and radius as $M=1.4~M_\odot$ and $R=$10 km. As already described, our cooling models assume the superburst energy $E_{\rm tot}$ is distributed uniformly over the later $P\le P_{\rm ign}$, and therefore is given as
\begin{eqnarray}
    E_{\rm tot} = \frac{E\dot{M}t_{\rm acc}}{(1+z)},
\end{eqnarray}
where $t_{\rm acc}=0.1~{\rm day}$ is the accretion episode prior to the onset of cooling and $\dot{M}=5\times10^{16}~{\rm g~s^{-1}}$ is the accretion rate during that pre-superburst phase, both of which are consistent with \cite{2025ApJ...986L..29A}. Thus, varying $E_{\rm tot}$ is the equivalent to varying three parameters of $E$, $\dot{M}$, and $t_{\rm acc}$.
\\
The final results on $P_{\rm ign}$ and $E_{\rm tot}$ are shown in Fig. \ref{fig:mcmcpe}. As we see, the presence of Urca leads to higher $P_{\rm ign}$ and slightly higher $E_{\rm tot}$. The most probable values (in the smallest $|\ln \mathcal{L}|$ case) are ($E_{\rm tot}$,$P_{\rm ign}$) = ($4.7\times10^{42}~{\rm erg}$, $2.1\times10^{26}~{\rm dyne~cm^{-2}}$) and ($3.5\times10^{42}~{\rm erg}$, $1.5\times10^{25}~{\rm dyne~cm^{-2}}$) in the presence and absence of Urca cooling, respectively. Observations of MAXI J1752$-$457 show $E_{\rm tot} = (5.3\pm0.7)\times10^{42}$
erg and $P_{\rm ign}=1.8\times10^{26}$ dyne cm$^{-2}$ \cite{2025ApJ...986L..29A}. Comparing our models with both of them, the presence of Urca is clearly favored. In particular, without Urca cooling, the ignition depth is far from the observations by an order of 1, while with Urca, the ignition depth is closer to the observations. This supports another piece of evidence for the presence of Urca pairs in the NS ocean.

\end{document}